\documentstyle[preprint,aps]{revtex}
\begin{document}
\title{SPIN FORCE DEPENDENCE OF NUCLEON STRUCTURE FUNCTIONS}
\author{H. J. Weber}
\address{Institute for Nuclear and Particle Physics\\
University of Virginia\\
Charlottesville, VA  22901, USA}
\maketitle
\begin{abstract}
Deep inelastic structure functions for the nucleon are
obtained in a constituent quark model on the light cone.
In the Bjorken limit the parton model is derived. Structure
functions from the hadronic tensor at the scale $\mu \sim
0.25$ GeV are evolved from (0.6 GeV$)^{2}$ to $-q^{2}\sim 15$
GeV$^{2}$. The model incorporates the quark boosts, is designed to
describe form factors and structure functions, and to provide a
link between the parton and constituent quark models. The main
effect of the spin force between quarks is described in terms of
smaller (and lighter) scalar $u-d$ quark pairs in the nucleon.
To account for the negative slope of $F^{n}_{2}(x)/F^{p}_{2}(x)$,
attraction between scalar $u-d$ quark pairs and incorporation of
boosts are required. The two-body spin force between quarks from
color magnetism also leads to a negative slope. The polarization
asymmetry $A^{p}_{1}$ is in fair agreement with the EMC data in
the valence quark region.  Revised hep-ph/9401342, Phys. Rev. D.
\end{abstract}

\section{Introduction}
At low momentum transfer $(-q^{2}<0.25$ GeV$^{2})$ the nonrelativistic
quark model (NQM$^{1})$ explains many of the nucleon's static properties
as originating from three valence quarks whose dynamics include a two-body
confinement potential and a two-body spin force motivated by one-gluon
exchange of perturbative quantum chromodynamics (QCD). The effective
 degrees of freedom at low energies are dressed or constituent quarks
which are expected to emerge in the spontaneous chiral symmetry breakdown
of QCD. Other degrees of freedom, such as gluons, are integrated out.
A chiral version of the NQM may be constructed by including various
soft-pion and electromagnetic amplitudes and those from quark current
 commutators.$^{2}$ In the NQM, estimates for the kinetic energies of
the $u$ and $d$ quarks are of the same order as the constituent quark
 mass, $m_{q}\sim m_{N}/3,$ leading to the conclusion that relativistic
 effects are important for these quarks. There are numerous contributions
in the literature that include relativistic corrections to order
$(v/c)^{2}$ in particle velocities compared to the velocity of light, or
$(p/m)^{2}$ in momentum/mass powers. However, for a nucleon matrix element
of the electromagnetic current, say, one obtains powers up to $(p/m)^{10}$.
To see this, let us describe each quark by a Dirac spinor with $S$-wave
upper and $P$-wave lower component in the static limit. Then the nucleon
spin wave function contains products of three such quark spinors and one
nucleon total-momentum spinor (see Eq. 2 below) that are coupled by Dirac
matrices to the nucleon spin. Altogether the quark current contributes
 up to two powers of momentum and each nucleon wave function up to four
giving up to ten powers for such current matrix elements. There is no a
priori reason to believe that relativistic corrections of lowest order up
to $(p/m)^{2}$ should dominate or that an expansion in $p/m$ powers would
converge rapidly.
\par
\medskip
A chief motivation for the development of the light-cone quark model is
to include relativistic effects to {\bf all} orders and avoid truncated
$p/m$ expansions. The model is formulated on the light cone for several
reasons. First, in Dirac's light-cone form of relativistic quantum
mechanics$^{3}$ one boost operator and two linear combinations of boost
and rotation generators are kinematic, i.e.independent of interactions,
which is crucial for the construction of form factors involving boosted
wave functions. On the other hand, two rotation generators become
interaction dependent so that rotational symmetry is more difficult to
implement. Second, deep inelastic structure functions are based on the
kinematics of the Bjorken limit $(q^{2}\rightarrow -\infty $ and
$P\cdot q\rightarrow \infty $,
with the scaling variable $x=-q^{2}/2P\cdot q$ finite), where the virtual
photon naturally probes the quark current matrix elements near the light
cone.
\par
\medskip
To stay as close as possible to the NQM, the light-cone quark model uses
the same parameters and nearly the same values of the NQM parameters.
The constituent quark mass $m_{q}\sim $ $m_{N}/3,$ and the proton
(quark core or confinement) radius is given by the inverse of the harmonic
oscillator constant, $\alpha $ $\sim m_{q}$, the main parameter of
the confinement potential.
\par
\medskip
The light-cone quark model$^{4}$ improves the magnetic form factor fits
to larger momentum transfers $-q^{2}\cong  1$ to 2 GeV$^{2}$, as well as
the $N\rightarrow \Delta (1232)$ $M1$-transition$^{5}$ and magnetic moments
of the baryon decuplet.$^{6}$
\par
\medskip
At high energy, though, the polarized EMC data$^{7}$ seem to imply that
the quarks contribute less than $15\%$ to the proton spin, as observed
in the singlet axialvector current matrix element. This is known as the
"spin crisis" caused by the EMC's polarized deep inelastic scattering (DIS)
data.$^{8}$ We are using the light-cone quark model to examine this
discrepancy and find that $A^{p}_{1}(x)$ can be described in the valence
quark region. This result is based on several ingredients.
First, the polarization asymmetry $A^{p}_{1}(x)$ and the ratio of structure
functions, $F^{n}_{2}(x)/F^{p}_{2}(x)$, depend sensitively on the spin
force between quarks. If there is attraction between $u-d$ quarks in the
nucleon, then $A^{p}_{1}(x)$ and the negative slope of
$F^{n}_{2}(x)/F^{p}_{2}(x)$ can be explained in the valence quark region,
provided boosts are built into the quark model. Second, both of these DIS
observables are ratios of structure functions which are only moderately
affected by uncertainties inherent in a perturbative evolution from low to
high momentum. Third, the vector sum of the constituent spins is not
Lorentz invariant. This point has recently been emphasized by
$Ma$ et al.$^{9}$ In the rest frame of the proton the spins of the
constituent quarks sum to the proton spin in the light-cone quark model.
In contrast the EMC data measure
$$\Delta q=\int {dx} [q^\uparrow (x)-q^\downarrow (x)],$$
where $q^\uparrow (x)$ and $q^\downarrow (x)$ are the probabilities of
finding a quark (or antiquark) of flavor $q$ with longitudinal momentum
fraction $x$ of the proton and polarization parallel and antiparallel to
the proton spin {\bf in the infinite momentum frame.} The quantity
$\Delta q$ is defined by the singlet axial current matrix element
$$<P,S|\bar{q}\gamma _{\mu }\gamma _{5}q|P,S>= \Delta qS_{\mu }$$
in a Lorentz invariant way. Thus $$\Delta \Sigma =\Delta u+\Delta d+\Delta s$$
is the sum of quark helicities in the infinite momentum frame, whereas
$$\Delta q=<M_{q}>\Delta q_{L}$$ differs from the spin sum in the proton rest
 frame, $\Delta q_{L}$, by the matrix element of an operator
$$M_{q}=[(p_{0}+p_{3}+m_{q})^{2}-\vec{p}^{2}_{T}]/[2(p_{0}+p_{3})
(p_{0}+m_{q})]$$
that depends on the transverse motion of the quarks.
\par
\medskip
Structure functions are usually described in the parton model whose
connection with the constituent quark model remains unclear. Our results
provide a link beween the parton model phenomenology and light-cone quark
models of the hadron spectroscopy.
\par
\medskip
The paper is organized as follows. The light-cone quark model and its
nucleon wave function are introduced in Section 2. The electromagnetic
form factors are presented in Section 3 and the DIS structure functions
in Section 4. The light-cone form of the nucleon wave function with the
 spin force from color magnetism is described in Section 5. Results are
discussed in Section 6.

\newpage
\section{ Light-Cone Quark Model With Spin Force }
The light-cone quark model$^{4,5,6}$ is based on Dirac's light-cone form
of relativistic many-body quantum mechanics,$^{3}$ where some boost
generators are kinematic, i.e. interaction independent. Free Melosh
rotations$^{10}$ are central in this model for the construction of
nonstatic three-quark wave functions for the nucleon (see Eqs.1,2 below).
Such a construction based on free quark spin states is also valid in
models where interactions in light-cone dynamics are added to the
three-quark mass (squared).$^{11}$ By comparison, bag models treat only
the interacting quark relativistically and violate translation invariance.
\par
A relativistic Gaussian wave function (see Eqs.3,7) is chosen to describe
the confined quark motion while staying as close as possible to the NQM
and using its parameters. More details are given in refs.4,5. To make the
model more realistic, we adopt a parameterization$^{12}$ that allows us to
simulate the effects of a spin force between quarks.
\par
The need for a spin interaction in the hadronic spectroscopy is known
and established.$^{1}$ Color magnetism$^{13}$ is attractive in spin 0
and repulsive in spin 1 quark pairs hence it splits the
$\Delta (1232)$-nucleon, $ \Sigma -\Lambda$ hyperon masses, etc.
and meets the desire for a simple explanation based on quantum
chromodynamics (QCD), the accepted gauge field theory of the strong
interaction.
\par
While in the nonrelativistic quark model (NQM) the strength $\alpha _{s}
\sim 1.6$ of the OGE is unrealistically large so that its spin-orbit
interaction actually spoils the success with the hadronic masses,$^{14}$
in a relativized CQM the OGE enters with a more realistic strength
$\alpha _{s} \sim 0.6$ and its spin orbit interaction is helpful in
the mass spectroscopy.$^{15}$
\par
The main effect of the spin interaction for the nucleon is a smaller
spatial size (and lighter mass) of scalar $u-d$ quark pairs, which can be
modeled by a larger transverse momentum spread $(\alpha _{>})$ in the
radial (relativistic harmonic oscillator) nucleon wave function $\phi _{N}$.
We adopt such a parameterization$^{12}$ in Eq.(3) for the internal proton
wave function
\begin{equation}
\psi _{N}=\phi _{N}(13,2) J_{N}(13,2)+\phi _{N}(23,1) J_{N}(23,1),
\label{1}
\end{equation}
\begin{eqnarray}
J_{N}(13,2)=\bar{u}(p_{1})(\gamma \cdot P+m_{N})\gamma _{5}C\bar{u}^{T}
(p_{3})[\bar{u}(p_{2})u_{N}(P)],\label{2}
\end{eqnarray}
\begin{eqnarray}
\phi _{N}(13,2) \sim \exp ((-1/6\alpha ^{2}_{>})[(m^{2}_{q}+
\nonumber \vec{q}^{2}_{2T})/x_{1}+(m^{2}_{q}+\vec{q}^{2}_{2T})/x_{3}]\\
-1/(6\alpha ^{2}_{<})[(m^{2}_{q}+\vec{Q}^{2}_{2T})/x_{2}
+\vec{Q}^{2}_{2T}/(1-x_{2})]),\label{3} \end{eqnarray}
where
$$\alpha ^{2}_{<}= \alpha ^{2}_{N}(1-D),  \alpha ^{2}_{>}=
\alpha ^{2}_{N}(1+D)$$
and $D$ is an adjustable deformation parameter. The
light-cone spinors denoted by $u(p_{i})$ or $u_{N}$ are solutions of the
free on-shell quark or nucleon Dirac equation with the metric conventions of
ref.16; they contain the light-cone boosts. $C = i\gamma ^{2}\gamma ^{0}$
is the charge conjugation matrix.
\par
The variable $q_{2}$ is the relative quark four-momentum between the up
quark (\#1) and down quark (\#3) and $Q_{2}$ between the up quark (\#2)
and the $ud$ pair $(\#'s  1 \& 3)$ so that
\begin{equation}
q_{2} = (x_{1}p_{3}-x_{3}p_{1})/(x_{1}+x_{3}),
\nonumber Q_{2} = (x_{1}+x_{3})p_{2}-x_{2}(p_{1}+p_{3})
\end{equation}
with $\sum_{i} x_{i} =1.$ The variables $q_{3}, Q_{3}$ and $q_{1},
Q_{1}$ are defined by cyclic permutation of the indices in Eq.4.
Equivalently, the
$$\vec{Q}_{iT}=\vec{k}_{iT}=\vec{p}_{iT}-x_{i}\vec{P}_{T}$$
are the internal quark momenta with
$\sum_{i} \vec{k}_{iT}=0.$ Transverse components (1,2) are denoted
by the subscript T.
The $\gamma _{5}$ matrix in the nonstatic spin wave function in Eq.(2) is
characteristic of a quark pair (1,3) coupled to spin 0 [and similarly a
quark pair (2,3) in the second term of Eq.(1)], while the
$\gamma \cdot P+m_{N}$ originates from the Melosh transformations of free
quarks to the light cone.$^{4,10}$
\newpage
\section { Electromagnetic Nucleon Form Factors }
The electromagnetic form factors of the nucleon are derived from the
matrix elements$^{4}$ of the $J^{+}$ current
\begin{eqnarray}
<N'|J^{+}|N> = \int {d\Gamma }\{\psi ^{\dag }_{N}\sum_{i}
\bar{u}(p'_{i})\gamma ^{+}(e_{i}/x_{i})u(p_{i})\psi_{N}\},
\label{5} \end{eqnarray}
in the Drell-Yan frame, where $q^{+}=0,$ so that
\begin{eqnarray}
\nonumber eF_{1}(q^{2})=m_{N}<N(P')\uparrow |J^{+}|N(P)\downarrow >/P^{+},\\
q_{L}eF_{2}(q^{2})/2m_{N}=-m_{N}<N(P')\uparrow |J^{+}|N(P)\downarrow >
/P^{+},\label{6} \end{eqnarray}
where $\vec{q}_{T}=(q_{1},q_{2})$ and $$q_{L}=q_{1}-iq_{2},q_{R}=q_{1}+iq_{2}.
$$ Substituting the nucleon wave function from Eqs.$1,2,3$
into Eq.5 we obtain
$$<N'|J^{+}|N> = \int {d\Gamma }\{2/3 \phi _{N}(23,1')\phi _{N}(23,1)$$
$$
\bar{u'}_{N}[p'_{1}](\gamma ^{+}/x_{1})[p_{1}]u_{N}Tr([P'][p_{2}][P][p_{3}])
$$
$$+2/3\phi _{N}(1'2,3)\phi _{N}(13,2)\bar{u'}_{N}[p_{2}]u_{N}
Tr([P'][p'_{1}](\gamma ^{+}/x_{1})[p_{1}][P][p_{3}])$$
$$+2/3\phi _{N}(23,1')\phi _{N}(13,2)\bar{u'}_{N}[p'_{1}]
(\gamma ^{+}/x_{1})[p_{1}][P][p_{3}][P'][p_{2}]u_{N}$$
$$+2/3\phi _{N}(1'3,2)\phi _{N}(23,1)\bar{u'}_{N}[p_{2}][P][p_{3}][P']
[p'_{1}](\gamma ^{+}/x_{1})[p_{1}]u_{N}+(1'\leftrightarrow 2',
2\leftrightarrow 1)$$
$$-1/3[\phi _{N}(23',1)\phi _{N}(23,1)\bar{u'}_{N}[p_{1}]u_{N}
Tr([P'][p_{2}][P][p_{3}](\gamma ^{+}/x_{3})[p'_{3}])$$
\begin{equation}
+\phi _{N}(23',1)\phi _{N}(13,2)\bar{u'}_{N}[p_{1}][P][p_{3}]
(\gamma ^{+}/x_{3})[p'_{3}][P'][p_{2}]u_{N}+(1\leftrightarrow 2)]\},
\eqnum{7} \end{equation}
where the primes denote the interacting quark in the final state with
momentum $p'_{i}$. We also abbreviate
$ \bar{u'}_{N}=\bar{u}_{N}(P')$, $u_{N}=u_{N}(P),$
$$[p_{i}]=(\gamma \cdot p_{i}+m_{q})/2m_{q},
[P']=(\gamma \cdot P'+m_{N})/2m_{N}$$ etc.
\par
The form of the first term in Eq.7 with $\phi $ from Eq.3 suggests using
$\vec{q}_{1T}$ and $\vec{Q}_{1T}$ as integration variables. The quark
momentum variables in terms of $q_{1},Q_{1}$ are given by
\begin{eqnarray}
\vec{k}_{1T} = \vec{p}_{1T}-x_{1}\vec{P}_{T} = \vec{Q}_{1T},
\vec{k}_{2T} = \vec{p}_{2T}-x_{2}\vec{P}_{T}
\nonumber =\vec{q}_{1T}-x_{2}\vec{Q}_{1T}/(1-x_{1}), \\
\vec{k}_{3T} = \vec{p}_{3T}-x_{3}\vec{P}_{T}
= -\vec{q}_{1T}-x_{3}\vec{Q}_{1T}/(1-x_{1}).\eqnum{8}
\end{eqnarray}
The term with $\phi (13,2')\phi (13,2)$ uses $\vec{q}_{2T}$ and
$\vec{Q}_{2T}$ as integration variables, while $\vec{q}_{3T}$ and
$\vec{Q}_{3T}$ are used in all cross terms such as $\phi (1'2,3)\phi (13,2)$.
Otherwise the analysis of the Dirac-spinor matrix elements and traces follows
that given by our work in ref.4. The perpendicular integrals over the
Gaussians are done analytically including the polynomial structure from the
Dirac $\gamma $-algebra, while the remaining integrals over $x_{1}$ and
$x_{2}$ are done numerically.
\newpage
\section { Structure Functions }
\par
\medskip
The deep inelastic form factors $W_{1,2}$ are defined in terms of the
Lorentz and gauge invariant expansion of the symmetric part of the
hadronic tensor
\par
\begin{equation}
W^{\mu \nu }=(-g^{\mu \nu }+q^{\mu }q^{\nu }/q^{2})W_{1}+W_{2}(P^{\mu }
-P\cdot qq^{\mu }/q^{2})(P^{\nu }-P\cdot qq^{\nu }/q^{2})/P^{2}.
\eqnum{9}
\end{equation}
The hadronic tensor derives from the imaginary part of the
forward virtual Compton scattering amplitude. It may be written in terms
of the quark current
$$ e\bar{u}(p')\gamma ^{\mu } u(p)/\sqrt {p'^{+} p^{+}}$$
as
\begin{eqnarray}
W^{\mu \nu }=(m^{2}_{q}/m_{N}) \int {d\Gamma}\psi ^{\dag }_{N}
\sum_{i} e^{2}_{i}/x_{i}\delta ((p_{i}+q)^{2}-m^{2}_{q})\bar{u}(p_{i})
\gamma ^{\mu }[p'_{i}]\gamma ^{\nu }u(p_{i})\psi _{N},
\eqnum{10}
\end{eqnarray}
where $$p'_{i}=p_{i}+q$$ holds for {\bf all four} momentum
components of the struck quark that becomes free in the Bjorken limit (see
Eq.15 below). The imaginary part of the energy denominator (together with
the denominator $p'^{+}$ of the currents) in light-cone perturbation
theory supplies the $\delta ((p_{i}+q)^{2}-m^{2}_{q})$. The transverse
(denoted by $T$) and + momentum components are conserved at the photon-quark
vertex. The invariant phase space volume element
\begin{equation}
d\Gamma  = (16\pi ^{3})^{-2}d^{2}q_{i}d^{2}Q_{i}dx_{1}dx_{2}dx_{3}\delta
(\sum_{j} x_{j}-1)/x_{1}x_{2}x_{3}\eqnum{11}
\end{equation}
reflects the separation of the internal and c.m. motion, as the
internal wave function $\psi _{N}(x_{i},q_{i},Q_{i},\lambda _{i})$ does
not change under kinematic Lorentz transformations and translations of the
nucleon. The relative momentum variables $q_{i}$ and $Q_{i}$ for quark
i=1,2,3 are defined as in Eq.(4) for quark 2, so that
$$
\vec{Q}_{iT}=\vec{k}_{iT}=\vec{p}_{iT}-x_{i}\vec{P}_{T},
\sum_{i} \vec{k}_{iT}=0.
$$
 Substituting the wave function from Eqs.1,2 we get more explicitly for
the proton
\begin{eqnarray}
W^{\mu \nu } = N \int {d\Gamma } \{\phi ^{2}_{N}(23,1)
[4\bar{u}_{N}[p_{1}]\gamma ^{\mu }[p'_{1}]\gamma ^{\nu }[p_{1}]u_{N}
\nonumber Tr([P][p_{2}][P][p_{3}])\\
+4\bar{u}_{N}[p_{1}]u_{N}Tr([P][p_{2}]\gamma ^{\mu }[p'_{2}]
\nonumber \gamma ^{\nu }[p_{2}][P][p_{3}])\\
+\bar{u}_{N}[p_{1}]u_{N}
\nonumber Tr([P][p_{2}][P][p_{3}]\gamma ^{\nu }[p'_{3}]\gamma ^{\mu
}[p_{3}])]\\
\nonumber +\phi ^{2}_{N}(13,2)[1\leftrightarrow 2]\\
+\phi _{N}(23,1)\phi _{N}(13,2)
[4\bar{u}_{N}[p_{1}]\gamma ^{\mu }[p'_{1}]\gamma ^{\nu }
\nonumber [p_{1}][P][p_{3}][P][p_{2}]u_{N}\\
+4\bar{u}_{N}[p_{1}][P][p_{3}][P][p_{2}]\gamma ^{\mu }[p'_{2}]
\nonumber \gamma ^{\nu }[p_{2}]u_{N}\\
+\bar{u}_{N}[p_{1}][P][p_{3}]\gamma ^{\nu }[p'_{3}]\gamma ^{\mu }
[p_{3}][P][p_{2}]u_{N} + 1\leftrightarrow 2]\}.\eqnum{12}
\end{eqnarray}
This expression may be further simplified using identities such as
\begin{equation}
[P][p_{i}][P]={1\over 2}(1+p_{i}\cdot P/m_{q}m_{N})[P],\eqnum{13}
\end{equation}
where, e.g.,
$$
m^{2}_{q}+x^{2}_{i}m^{2}_{N}-2x_{i}p_{i}\cdot P = Q^{2}_{i}
=-\vec{Q}^{2}_{iT}=(p_{i}-x_{i}P)^{2},$$
$$
m^{2}_{q}+x^{2}_{1}m^{2}_{N}-2x_{1}p_{1}\cdot P = [q_{3}-x_{1}Q_{3}
/(1-x_{3})]^{2},$$

\begin{equation}
m^{2}_{q}+x^{2}_{2}m^{2}_{N}-2x_{2}p_{2}\cdot P = [-q_{3}-x_{2}Q_{3}
/(1-x_{3})]^{2}.\eqnum{13'}
\end{equation}
Since $$p'_{i}=p_{i}+q$$ holds for all four components, it is easy to
verify that $W^{\mu \nu }$ of Eq.10 is gauge invariant:
$$W^{\mu \nu }q_{\nu }=0=q_{\mu }W^{\mu \nu }$$ involve the typical
terms of the form
$$
(\gamma \cdot p_{i}+m_{q})\gamma \cdot q(\gamma \cdot p'_{i}+m_{q})=
(\gamma \cdot p_{i}+m_{q})(\gamma \cdot p'_{i}-\gamma \cdot p_{i})
 (\gamma \cdot p'_{i}+m_{q})$$
$$=(m_{q}+\gamma \cdot p_{i})(m_{q}-m_{q})(m_{q}+\gamma \cdot p'_{i})=0,$$
$$(\gamma \cdot p'_{i}+m_{q})\gamma \cdot q(\gamma \cdot p_{i}+m_{q})
=(\gamma \cdot p'_{i}+m_{q})(\gamma \cdot p'_{i}
 -\gamma \cdot p_{i})(\gamma \cdot p_{i}+m_{q})$$
\begin{equation}
=(m_{q}+\gamma \cdot p'_{i})(m_{q}-m_{q})(m_{q}+\gamma \cdot p_{i})=0,
\eqnum{14}
\end{equation}
using
$$
(\gamma \cdot p_{i})^{2}=p^{2}_{i}=m^{2}_{q}=(\gamma \cdot p'_{i})^{2}
={p'}_{i}^{2}.$$
\par
\medskip
We define $F_{1}(x,q^{2})= m_{N}W_{1}$ and $F_{2}(x,q^{2})= \nu W_{2}$
with the energy transfer $m_{N}\nu  = P\cdot q\rightarrow \infty $, while
the longitudinal Bjorken-Feynman momentum fraction $x=-q^{2}/(2m_{N}\nu )$
stays finite in the approach to scaling as $q^{2}\rightarrow -\infty $ and
$\nu \rightarrow +\infty $. In the Bjorken limit
$$
2P\cdot q=2m_{N}\nu \rightarrow \infty , q^{2}\rightarrow -\infty ,
0< x=-q^{2}/2P\cdot q <1,
$$
\begin{equation}
F_{1}(x,q^{2})=m_{N}W_{1}(q^{2},\nu )\rightarrow F_{1}(x), F_{2}(x,q^{2})=
\nu W_{2}(q^{2},\nu )\rightarrow F_{2}(x).
\eqnum{15}
\end{equation}
In light-cone variables it is convenient to work in a frame where the
nucleon moves along the $z$ (or 3-)axis: $P^{\mu } =(P^{+}>0,
P^{-}=m^{2}_{N}/P^{+},\vec{P}_{T}=0)$. (In the nucleon rest frame
$P^{+}=m_{N}$ and the photon energy $q_{0}=\nu \rightarrow \infty $
from Eq.15.) If we take $q^{z}<0,$ then
$$
q^{-}=q_{0}-q^{z}=-2xm_{N}\nu /q^{+}\sim 2m_{N}\nu /P^{+}\rightarrow \infty ,
$$
while $$q^{+}=q_{0}+q^{z}\sim -xP^{+}$$ and $\vec{q}_{T}$ stay
finite. In fact, from Eq.15 we obtain
$$
2m_{N}\nu (1+xP^{+}/q^{+})=m^{2}_{N}q^{+}/P^{+},
$$
so that
\begin{equation}
q^{+}/P^{+}=\nu /m_{N}\{1-[1+2xm_{N}/\nu ]^{1/2}\}=-\xi \sim -x,
\eqnum{16}
\end{equation}
as $\nu \rightarrow \infty $,
where $\xi $ is the Nachtmann variable that may also be written as
$$
\xi = 2x/\{1+[1+2xm_{N}/\nu ]^{1/2}\}
=-q^{2}/\{\nu +[\nu ^{2}-q^{2}]^{1/2}\}m_{N}.
$$
If we use $p'^{-}_{i}=p^{-}_{i}+q^{-}$ for the {\bf interacting}
quark, then
$$
2P\cdot q=m^{2}_{N}q^{+}/P^{+}+P^{+}\{m^{2}_{q}+
(\vec{p}_{iT}+\vec{q}_{T})^{2}]/(x_{i}P^{+}+q^{+})
-(m^{2}_{q}+\vec{p}^{2}_{iT})/x_{i}P^{+}\}\sim 2m_{N}\nu
$$
requires that ($q^{+}/P^{+}=-\xi \rightarrow -x_{i},$or) $x_{i}
\rightarrow x$ in the Bjorken limit. (Note that $q^{+}<0$ is appropriate
for bosons; and the virtual photon
is far off its mass shell with $q^{2}\sim q^{+}q^{-}\sim -2xm_{N}\nu
\rightarrow -\infty $.)
Equivalently, the delta function in Eq.10 may be rewritten as
$$
\delta ((p_{i}+q)^{2}-m^{2}_{q})=\delta (q^{2}+2p_{i}\cdot q)
=\delta (q^{-}(q^{+}+p^{+}_{i})+p^{-}_{i}q^{+}-2\vec{p}_{iT}\cdot
\vec{q}_{T}-\vec{q}^{2}_{T})$$
\begin{eqnarray}
=\delta (P^{+}q^{-}(x_{i}-\xi )-p^{-}_{i}\xi P^{+}-2\vec{p}_{iT}\cdot
\vec{q}_{T}-\vec{q}^{2}_{T})\rightarrow -\xi \delta (x_{i}-\xi )/q^{2}
\eqnum{17}
\end{eqnarray}
in the Bjorken limit, where also $\xi \rightarrow x$.
\par
We are now ready to project
\begin{equation}
W_{1}=P^{1}_{\mu \nu }W^{\mu \nu }, W_{2}=P^{2}_{\mu \nu }W^{\mu \nu },
\eqnum{18}
\end{equation}
from $W^{\mu \nu }$, where
$$
P^{1}_{\mu \nu }=\{-g_{\mu \nu }+P_{\mu }P_{\nu }
/(1-\nu ^{2}/q^{2})m^{2}_{N}\}/2,
$$
\begin{equation}
P^{2}_{\mu \nu }=\{-g_{\mu \nu }+3P_{\mu }P_{\nu }/(1-\nu ^{2}/q^{2})
m^{2}_{N}\}/2(1-\nu ^{2}/q^{2}).\eqnum{19}
\end{equation}
Using
$$
\gamma _{\mu }[p'_{i}]\gamma ^{\mu } = 3-2[p'_{i}]
$$
in Eq.12 the typical term in $W^{\mu \nu }g_{\mu \nu }$ for the
proton scales as
\begin{equation}
[p_{i}][p'_{i}][p_{i}]/q^{2} = (m^{2}_{q}+p_{i}\cdot p'_{i})
[p_{i}]/(2m^{2}_{q}q^{2}) \sim -[p_{i}]/4m^{2}_{q}\eqnum{20}
\end{equation}
including $q^{-2}$ from the delta function in Eq.17 and using
$$
p_{i}\cdot p'_{i}=m^{2}_{q}-q^{2}/2\sim m_{N}\nu x.
$$
Replacing effectively in $W^{\mu \nu }g_{\mu \nu }$ of Eq.12 all the terms
$[p_{i}]\gamma ^{\mu }[p'_{i}]\gamma _{\mu }[p_{i}]$ by
$[p_{i}]=(\gamma \cdot p_{i}+m_{q})/2m_{q}$ generates in $W_{1,2}$ the
spin structure of $\psi ^{\dag }_{N}\delta (x_{i}-x)\psi _{N}$.
\par
\medskip
All terms $\sim  P_{\mu }W^{\mu \nu }P_{\nu }$ in $P^{1,2}_{\mu \nu }
W^{\mu \nu }$ scale to zero because of the extra $\nu \rightarrow \infty $
in the denominator of the $P_{\mu }P_{\nu }$ terms in $P^{1,2}_{\mu \nu }$.
Substituting Eqs.17,20 into Eq.12, we obtain the structure function of the
parton model
\begin{equation}
F_{1}(x)=\int {d\Gamma} \psi ^{\dag }_{N}\Sigma e^{2}_{i}\delta
(x_{i}-x)\psi _{N}/2=\sum_{i} e^{2}_{i}q_{i}(x)/2,
\eqnum{21}
\end{equation}
where the quark probabilities $q_{i}$ are derived from the
light-cone quark model wave function $\psi _{N}$:
\begin{equation}
q_{\lambda _{i}}(x)=\sum_{\lambda _{j},j \neq i}(16\pi ^{3})^{-2}
\int {[dx][d^{2}{\bf k}_{jT}]}\delta (x_{i}-x)
|\psi _{N}(x_{j},{\bf k}_{jT},\lambda _{j})|^{2}.
\eqnum{22}
\end{equation}
For $F_{2}$ the extra factor $\nu /(1-\nu ^{2}/q^{2}) \rightarrow 2xm_{N}$,
so that $F_{2}(x)=2xF_{1}(x)$ is obtained. From the normalization of the
nucleon wave function $\psi _{N}$ we obtain the sum rule for $F_{2}$:
\begin{equation}
\int {dx}F_{2}(x)/x = \sum_{i} e^{2}_{i}.\eqnum{23}
\end{equation}
The polarized structure functions $g_{i}(x,q^{2})$ are defined by the
antisymmetric part of the hadronic tensor
\begin{equation}
W^{\mu \nu }_{A} = \epsilon ^{\mu \nu \sigma \rho } q_{\sigma }
[S_{\rho }g_{1}(x,q^{2})+[S_{\rho }-P_{\rho }q\cdot S/(m_{N}\nu )]
g_{2}(x,q^{2})],\eqnum{24}
\end{equation}
with the spin vector
\begin{equation}
S_{\rho } = \bar{u}_{N}\gamma _{\rho }\gamma _{5}u_{N}.
\eqnum{25}
\end{equation}
In order to extract $g_{1}$ and $g_{2}$, we expand the relevant terms
\begin{eqnarray}
\gamma ^{\mu }\gamma \cdot p'_{i}\gamma ^{\nu } =
({p'}_{i}^{\mu }\gamma ^{\nu }+{p'}_{i}^{\nu }
\gamma ^{\mu }-g^{\mu \nu }\gamma \cdot p'_{i})+
i\epsilon ^{\mu \nu \sigma \rho }p'_{i\sigma }
\gamma _{\rho }\gamma _{5}
\eqnum{26}
\end{eqnarray}
in $W^{\mu \nu }$ in Eq.10. The first three symmetric terms in Eq.26 have
been included in $W^{\mu \nu }_{S}$ in Eq.12, while the last antisymmetric
term generates $W^{\mu \nu }_{A}$. Omitting the
$i\epsilon ^{\mu \nu \sigma \rho }$ factor and noticing that
$p'_{i}=p_{i}+q$ provides the only $q$-dependence, we obtain
\begin{equation}
g_{2}(x)\equiv 0,\eqnum{27}
\end{equation}
and
\begin{eqnarray}
\nonumber g_{1}(x,q^{2})S_{\rho }=\nu (m_{q}/m_{N})^{2}\int {d\Gamma }
\psi ^{\dag }_{N}\sum_{i} e^{2}_{i}//\delta ((p_{i}+q)^{2}-m^{2}_{q})
\bar{u}_{i}\gamma _{\rho }\gamma _{5}u_{i}\psi _{N}/x_{i}.
\eqnum{28}
\end{eqnarray}
In the Bjorken limit Eq.17 yields for the delta function
\begin{equation}
\nu \delta ((p_{i}+q)^{2}-m^{2}_{q})\rightarrow -x\nu \delta (\xi -x_{i})/q^{2}
\rightarrow \delta (x-x_{i})/2m_{N}.\eqnum{29}
\end{equation}
The $q^{2}$ factor in Eq.29 is the only $q$-dependence in Eq.28.
 Substituting Eq.29 into Eq.28 along with the nucleon wave function
$\psi _{N}$ from Eqs.1,2,3 we obtain
\begin{equation}
\nonumber g_{1}(x)S_{\rho } = (1/2)(m_{q}/m_{N})^{2}\int {d\Gamma }\\
\{\phi ^{2}(23,1)[23,1]_{\rho }+\phi ^{2}(13,2)[13,2]_{\rho }
+\phi (23,1)\phi (13,2)[I]_{\rho }\},
\eqnum{30}
\end{equation}
where
$$
[23,1]_{\rho } = 4\bar{u}_{N}[p_{1}]\gamma _{\rho }\gamma _{5}[p_{1}]u_{N}
Tr([P][p_{2}][P][p_{3}])$$
$$+4\bar{u}_{N}[p_{1}]u_{N}Tr([P][p_{2}]\gamma _{\rho }\gamma _{5}[p_{2}][P]
[p_{3}])$$
$$-\bar{u}_{N}[p_{1}]u_{N}Tr([P][p_{2}][P][p_{3}]\gamma _{\rho }\gamma _{5}
[p_{3}]),$$
$$
[13,2]_{\rho } = [23,1]_{\rho } ( 1\leftrightarrow 2),
$$
$$
[I]_{\rho } = 4\bar{u}_{N}[p_{1}]\gamma _{\rho }\gamma _{5}[p_{1}][P][p_{3}]
[P][p_{2}]u_{N}$$
$$+4\bar{u}_{N}[p_{1}][P][p_{3}][P][p_{2}]\gamma _{\rho }\gamma _{5}
[p_{2}]u_{N}$$
\begin{equation}
-\bar{u}_{N}[p_{1}][P][p_{3}]\gamma _{\rho }\gamma _{5}[p_{3}][P][p_{2}]u_{N}\\
+(1\leftrightarrow 2).\eqnum{31}
\end{equation}
We analyze Eq.31 using the identities in Eqs.$13,13'$ and
\begin{equation}
[p_{i}]\gamma _{\rho }\gamma _{5}[p_{i}]
= (\gamma _{\rho }+p_{i\rho }/m_{q})\gamma _{5}[p_{i}],
\eqnum{32}
\end{equation}
\begin{equation}
[P][p_{i}]u_{N} = (1/2)(1+p_{i}\cdot P/m_{q}m_{N})u_{N}.
\eqnum{32'}
\end{equation}
This yields for the first term of $[23,1]_{\rho }$, for example,
$$
4\bar{u}_{N}[p_{1}]\gamma _{\rho }\gamma _{5}[p_{1}]u_{N}Tr([P][p_{2}][P]
[p_{3}])$$
\begin{eqnarray}
= 2(1+p_{2}\cdot P/m_{q}m_{N})(1+p_{3}\cdot P/m_{q}m_{N})\bar{u}_{N}
(\gamma _{\rho }+p_{1\rho }/m_{q})\gamma _{5}[p_{1}]u_{N}.
\eqnum{33}
\end{eqnarray}
Integrating Eq.33 over the internal momentum coordinates yields the
expression
\begin{eqnarray}
(\bar{u}_{N}\gamma _{\rho }\gamma _{5}u_{N})\int {d\Gamma }\phi ^{2}
(23,1)(1+p_{2}\cdot P/m_{q}m_{N})(1+p_{3}\cdot P/m_{q}m_{N})e^{2}_{1}/x_{1},
\eqnum{34}
\end{eqnarray}
which, except for the factor $S_{\rho }$, is precisely the
corresponding term in the quark probability in Eq.22. The trace in the
second term of $[23,1]_{\rho }$,
\begin{eqnarray}
4\bar{u}_{N}[p_{1}]u_{N}Tr([P][p_{2}]\gamma _{\rho }\gamma _{5}
[p_{2}][P][p_{3}]) = (1+p_{3}\cdot P/m_{q}m_{N})\bar{u}_{N}[p_{1}]u_{N}
Tr((\gamma _{\rho }+p_{2\rho }/m_{q})\gamma _{5}[p_{2}][P])
\eqnum{35}
\end{eqnarray}
vanishes obviously, and that of the third term similarly, etc.
For the first term in $[I]_{\rho }$ we obtain
\begin{eqnarray}
4\bar{u}_{N}[p_{1}]\gamma _{\rho }\gamma _{5}[p_{1}][P][p_{3}][P][p_{2}]u_{N}
= (1+p_{2}\cdot P/m_{q}m_{N})(1+p_{3}\cdot P/m_{q}m_{N})\bar{u}_{N}
(\gamma _{\rho }+p_{1\rho }/m_{q})\gamma _{5}[p_{1}]u_{N}
\eqnum{36}
\end{eqnarray}
which, on integrating over the internal momentum variables, yields
the contribution
$$
(\bar{u}_{N}\gamma _{\rho }\gamma _{5}u_{N})\int {d\Gamma }\phi (23,1)
\phi (13,2)(1+p_{2}\cdot P/m_{q}m_{N})(1+p_{3}\cdot P/m_{q}m_{N})
e^{2}_{1}/x_{1}$$
to $g_{1}S_{\rho }$, and similar ones for all other terms in
$[I]_{\rho }$ in Eq.31. On comparing with the individual terms in the quark
probabilities in Eq.22 we find that these contributions in Eqs.34,35,36 and
so on generate the longitudinal polarized structure function
\begin{equation}
g_{1}(x) = {1\over 2}\sum_{i} e^{2}_{i}[q^\uparrow (x)-q^\downarrow (x)]
\eqnum{37}
\end{equation}
of the parton model.
\par
\medskip
While the {\bf interacting} quark is treated as a current quark in the
parton model, we have obtained the same form of the structure functions
using the quark currents of the constituent quark model. In ref.18 such
structure function results are taken to represent the non-perturbative
input at a low energy scale $\mu \sim 0.25$ GeV and are then evolved
to high $q^{2}$. A perturbative QCD evolution from such a low resolution
 value to high $q^{2}$ is unreliable. To avoid this problem we extract
the momentum dependent structure functions $F_{i}(x,q^{2})$ from the
hadronic tensor at the $\mu =0.25$ GeV scale (before the Bjorken limit).
Then we evolve them from $-q^{2}=(0.6$ GeV$)^{2}$, where the relativistic
quark model is clearly valid, to the $-q^{2}\sim 15$ GeV$^{2}$ of the
EMC data.
\newpage
\section{ Light-Cone Quark Model With Color Magnetism }
It is well known that a nucleon wave function based on the symmetric
$[56,0^{+}]$ representation of $SU(6)$ generates deep inelastic structure
functions in disagreement with the data. In particular, the ratio of
neutron-to-proton structure functions $F^{n}_{2}(x)/F^{p}_{2}(x)=2/3$ is
constant in contrast to the negative slope of the data shown in Fig.2.
\par
In the NQM the $SU(6)$ symmetry is broken by the two-body spin interaction
from color magnetism besides the constituent quark masses $m_{u}=m_{d}
=m_{q}\neq m_{s}$. Its most important effect is the admixture of the
$[70,0^{+}] SU(6)$ configuration to the $[56,0^{+}]$. In the NQM a
truncated wave function may be written in coordinate space as
\begin{equation}
|N> = a_{56}|N,56>+a_{70}|N,70>\eqnum{38}
\end{equation}
with $a_{56}=0.95$ and $a_{70}\sim 0.2$ for $m_{q}=0.33$ GeV
and $\alpha =0.32$ GeV. (Note that $a_{56}=a_{S}$ and $a_{70}=-a_{S_{M}}$
in refs.17, 18 and $|N,70>\rightarrow -|^{2}S_{M}>$ in the static limit.)
All other configurations have substantially smaller admixture coefficients
except for the $2S=S'$ state which, if included, would renormalize the $1S$
 radial wave function of the $[56,0^{+}]$ and lower the
$F^{n}_{2}(x)/F^{p}_{2}(x)$ slope. In order to evaluate deep inelastic
structure functions for the $N$ wave function in Eq.(38) we translate it to
 the light cone in momentum space. The Gaussian radial $S$-wave function
\begin{equation}
\phi _{S} = \exp [-\alpha ^{2}(\vec{\rho }^{2}+\vec{\lambda }^{2})/2]
\rightarrow \phi _{0} = \exp [-\sum_{i=1}^3 M_{j}/6\alpha ^{2}],\eqnum{39}
\end{equation}
where
\begin{equation}
\vec{\rho } = (\vec{r}_{1}-\vec{r}_{2})/\sqrt{2}, \vec{\lambda }
= (\vec{r}_{1}+\vec{r}_{2}-2\vec{r}_{2})/\sqrt{6}
\eqnum{40}
\end{equation}
are the usual relative quark variables in coordinate space and
$$
M_{j} = (\vec{k}^{2}_{jT}+m^{2}_{q})/x_{j},
$$
\begin{equation}
\sum_{j} M_{j}= \vec{q}^{2}_{3T}(1-x_{3})/x_{1}x_{2}+\vec{Q}^{2}_{3T}/x_{3}
(1-x_{3})+\sum_{j} m^{2}_{q}/x_{j}.\eqnum{41}
\end{equation}
in light cone variables.
\par
\medskip
The $[70,0^{+}]$ wave function of mixed symmetry involves the radial wave
functions
\begin{equation}
\phi ^{\lambda } = (\vec{\rho }^{2}-\vec{\lambda }^{2})\phi _{S} ,
\phi ^{\rho } = 2\vec{\rho }\cdot \vec{\lambda }\phi _{S},
\eqnum{42}
\end{equation}
which are translated to the light cone using the Fourier transform
\begin{equation}
\int {d^{3}}\rho \exp (i\vec{p}_{\rho }\cdot \vec{\rho }
-\alpha ^{2}\vec{\rho }^{2}/2)\vec{\rho }^{2}
=(2\pi /\alpha ^{2})^{3/2}\{3/\alpha ^{2}-\vec{p}^{2}_{\rho }
/\alpha ^{4}\}\exp (-\vec{p}^{2}_{\rho }/2\alpha ^{2})
\eqnum{43}
\end{equation}
and the corresponding ones for $\vec{\lambda }$ and its conjugate
momentum variable $\vec{p}_{\lambda }$ and $\vec{\rho }\cdot \vec{\lambda }$.
Note the important relative minus sign in the Fourier transforms
\begin{equation}
\vec{\rho }^{2}-\vec{\lambda }^{2} \rightarrow \vec{p}^{2}_{\lambda }
-\vec{p}^{2}_{\rho } , \vec{\rho }\cdot \vec{\lambda }
\rightarrow -\vec{p}_{\rho }\cdot \vec{p}_{\lambda }.
\eqnum{44}
\end{equation}
In the nonrelativistic limit the longitudinal momentum fractions
$x_{j}\rightarrow 1/3,$ and
\begin{equation}
3(M_{1}+M_{2}-2M_{3}) \rightarrow \vec{p}^{2}_{\rho }-\vec{p}^{2}_{\lambda },\\
M_{2}-M_{1} \rightarrow -2\sqrt{3}\vec{p}_{\rho }\cdot \vec{p}_{\lambda },\\
\sum_{j} M_{j}-9m^{2}_{q}\rightarrow  3(\vec{p}^{2}_{\rho }
+\vec{p}^{2}_{\lambda }).
\eqnum{45}
\end{equation}
The translation of the spin-isospin wave functions to the
uds-basis on the light cone is discussed in detail in ref.4. Altogether
the $[70,0^{+}]$ wave function on the light cone takes the form
$$
\psi _{N} = a_{56}\psi _{56} - a_{70}\psi _{70}
$$
$$
\psi _{56} = \phi _{0}N_{0}\{J_{N}(13,2)+J_{N}(23,1)\}
$$
\begin{eqnarray}
\psi _{70}=N_{\lambda }\phi _{0}(M_{1}+M_{2}-2M_{3}) [J_{N}(13,2)
+J_{N}(23,1)] /\alpha ^{2}
+N_{\rho }\phi _{0}(M_{2}-M_{1})J_{N}(12,3)/\alpha ^{2}
\eqnum{46}
\end{eqnarray}
with $J_{N}$ of Eq.(2). The positive constants $N_{\rho }$ and $N_{\lambda }$
normalize each of the mutually orthogonal terms in $\psi _{70}$ of Eq.(46).
\medskip The structure function $F_{2}(x)$ can
be calculated from the parton model formulas, Eqs.21,22, which are
derived along lines similar to Section 4.
\newpage
\section{ Discussion of Results and Conclusion }
Our numerical results for the electromagnetic form factors (based on
pointlike constituent quarks with a mass $m_{q}\sim m_{N}/3,$
without anomalous magnetic moments and axial vector quark coupling
constant $g^{q}_{A}=1$ at $q^{2}=0)$ reproduce the static properties of
the proton and neutron in ref.4 for $D=0,$ providing a test of our
numerical and symbolic codes. The deformation parameter $D=0.37$ is
obtained from the ratio of neutron to proton structure functions
$F^{n}_{2}(x)/F^{p}_{2}(x)$. This deformation corresponds to
attraction between scalar $u-d$ quark pairs in the nucleon and is
causing relatively minor changes in the electromagnetic form factors
of the nucleon which are shown in Figs.1,2. E.g., for $m_{q} = 0.33$ GeV$
/c^{2}$, we obtain the nucleon magnetic moments $\mu _{p}=2.475$ n.m.
and $\mu _{n}=-1.58$ n.m. which increase in absolute value by $5\%$ to
$15\%$ when pion cloud corrections are included.$^{19}$
\par
\medskip
For $D=0,$ we also obtain the constant $SU(6)$ value 2/3 for
$F^{n}_{2}(x)/F^{p}_{2}(x)$. For $D\neq 0$ the $SU(6)$ symmetry is broken
and $F^{n}_{2}(x)/F^{p}_{2}(x)$ is no longer constant, but falls off with $x$
 increasing towards 1, as shown by the solid line in Fig. 3, provided
there is attraction between scalar $u-d$ quark pairs $(D=+0.37)$. For
repulsion, $D=-0.37, a$ positive slope is obtained (dot-dashed line in Fig.3).
 This feature also shows up in the results for the mixed [56]-[70] wave
function of Section 5 which incorporates approximately the main effect of
the spin force from color magnetism (dotted line in Fig.3). Reversing the
phase of the admixture coefficient $a_{70}$ yields a positive slope
(dashed line in Fig.3). The sensitivity of this ratio of structure
functions for $x>1/4$ to the spin force is remarkable. Thus, the slope of
$F^{n}_{2}(x)/F^{p}_{2}(x)$ for $x>1/4$ is a sensitive probe of the spin
interaction between quarks.
\par
\medskip
These structure functions have been calculated at the low energy scale
$\mu $ in Section 4, where the nucleon is taken to consist of constituent
valence quarks only, while sea quarks and gluons are neglected. The results
 at high $q^{2}$ are generated radiatively by means of a perturbative QCD
 evolution which depends on the running coupling constant $\alpha _{s}$
that contains the renormalization group scale parameter
$\Lambda _{QCD}$ in logarithmic form
$\ln (-q^{2}/\Lambda ^{2}_{QCD})$. The results from the evolution
do not depend on the scale $\mu $ or $\Lambda _{QCD}$ separately,
but on the logarithmic ratio $L=\ln (-q^{2}/
\Lambda ^{2}_{QCD})/\ln (\mu ^{2}/\Lambda ^{2}_{QCD})$.
Based on the value $\Lambda _{QCD}\sim 0.2$ GeV, $L= 14.7$
has been extracted$^{18}$ from the second moment $<F^{N}_{2}(-q^{2}=15$
GeV$^{2})>_{2}=0.127$ of the (average) nucleon structure function from the
 EMC data for the deuteron. This value of $L$ yields the low energy scale
$\mu \sim 0.25$ GeV,$^{18} a$ value that is consistent with
$\alpha \sim m_{q}\sim m_{N}/3,$ but so low that a perturbative
evolution gives rise to large changes of quark distributions and structure
 functions and is not trustworthy. Nonetheless it is interesting to note
that our results for $F^{p}_{2}(x)$ in Fig.4 and $xu_{v}(x)$ in Fig.5 and
those in ref.18 are similar. Our results for the valence quark probability
$u_{v}(x)$ are shown (as solid line for the attractive spin force,
$D=0.37,$ and dashed line for the mixed [56]-[70] color hyperfine wave
function) in Fig.5 and are to be compared with the dot-dashed curve from
(Fig.3 in) ref.18 corresponding to the Isgur-Karl quark model. The smaller
dot-dashed curve is their result of an evolution from (0.25 GeV$)^{2}$ to
15 GeV$^{2}$. (Note that the perturbative evolution is known to become
 unreliable near the endpoints $x=0$ and 1.) While these evolution effects
are large for $xu_{v}(x)$, they are much smaller, though not negligible,
for the ratio $F^{n}_{2}(x)/F^{p}_{2}(x)$ for $x>1/4.$ Hence light-cone
quark model results for {\bf ratios} of structure functions are more
reliable because of their much smaller modifications due to the QCD
evolution. In order to avoid the evolution at too low $q^{2}$, we have
extracted the structure functions $F_{i}(x,q^{2})$ from the hadronic
 tensor and evolved them from $-q^{2}=(0.6$ GeV$)^{2}$, where our
relativistic quark model is still valid, but $F_{2}(x,q^{2})$
$\neq 2xF_{1}(x,q^{2})$. At $-q^{2}=15$ GeV$^{2}$ we have verified that
$F_{2}(x,q^{2})=2xF_{1}(x,q^{2})\sim F_{2}(x)$. These results are
shown in the dot-dashed and short-dashed lines in Fig.4. From the latter
we now see more clearly the missing sea-quarks at $x<1/3.$ Note that the
shift of the $F^{p}_{2}$ peak to smaller $x$ mainly comes from the QCD
evolution. Choosing a higher value than (0.6 GeV$)^{2}$ makes the QCD
corrections too small. However, including sea quarks that mainly
contribute at small $x$ would effectively shift the $F^{p}_{2}(x)$
peak to lower $x$ values, allow us to raise the lower limit (0.6 GeV$)^{2}$
 of perturbative QCD corrections and extend the validity of the model to
lower x.
\par
\medskip
The different slope results in Fig.6 (see also ref.24 for a discussion of
earlier work) show that $F^{n}_{2}(x)/F^{p}_{2}(x)$ is also sensitive to
the boosts built into the Melosh transformations, which are missing in the
nonrelativistic Karl-Isgur wave function used in ref.18. (Note also that
in the relativized NQM versions of ref.15 the relevant $[70,0^{+}]$
admixture coefficient $a_{70}$ does not change by much.)
\par
Our result for the polarization asymmetry $A^{p}_{1}\simeq 2xg^{p}_{1}(x)$/
$F^{p}_{2}(x)$ (the solid line in Fig.7) is also in fair agreement with
the EMC data in the valence quark region. (A perturbative QCD evolution of
the solid line is shown in Fig.7 as dot-dashed line which we consider as
unreliable. The data are nearly $q^{2}$ independent.)
\par
\medskip
In summary, the polarization asymmetry of the proton and the ratio of
neutron to proton structure functions are sensitive probes of the spin
force between quarks. Such ratios of structure functions are observables
that are only moderately affected by uncertainties involved in a perturbative
evolution to high momentum. Electromagnetic nucleon form factors and ratios
of deep inelastic structure functions are compatible in a constituent
light-cone quark model with an attractive spin force.
\par
\medskip
\section*{ Acknowledgements }
It is a pleasure to thank Xiaotong Song for stimulating discussions and his
help in providing the evolved results based on ref.22. This work was
supported in part by the U.S. National Science Foundation.
\newpage
\begin{figure}
\caption{
Magnetic proton form factor normalized to the dipole shape
$(1-q^{2}/m^{2}_{D})^{2}$ for $m^{2}_{D}=0.71$ GeV$^{2}$. The solid line is
 our light-cone quark model with $\alpha =0.35$ GeV, $m_{q}=0.33$ GeV,
 $D=0.37$ with attractive spin force, the dashed line is for $D=0,^{4}$
no spin force. The dot-dashed line represents the nonrelativistic
constituent quark model (NQM). The experimental data are from ref.20.}
\label{fig1}
\end{figure}
\begin{figure}
\caption{
Proton charge form factor normalized to the dipole shape. The curves
are denoted as in Fig.1}
\label{fig2}
\end{figure}
\begin{figure}
\caption{
Ratio of unpolarized neutron to proton structure functions. The solid line
is for attraction in scalar $u-d$ quark pairs in the light-cone quark model
with $D=0.37.$ The dot-dashed line is for repulsion, $D=-0.37.$ The dotted
line is for the spin force from color magnetism and the dashed line is with
opposite sign of $a_{70}(=-0.2)$. Data are from ref.21.}
\label{fig3}
\end{figure}
\begin{figure}
\caption{
Unpolarized proton structure function $F^{p}_{2}(x)$. The upper solid line
is for the attractive spin force, $D=0.37,$ in the light-cone quark model;
the lower solid line is its evolution from (0.25 GeV$)^{2}$ to 15 GeV$^{2}$
based on ref.22. The upper dashed line is for the spin force from color
 magnetism. The dot-dashed line is $F^{p}_{2}(x,q^{2})$ at
$-q^{2}=(0.6$ GeV$)^{2}$ from the hadronic tensor at the $\mu =0.25$ GeV
scale of the light-cone quark model; the short-dashed line is its evolution
from (0.6 GeV$)^{2}$ to 15 GeV$^{2}$. Data are from ref.23.}
\label{fig4}
\end{figure}
\begin{figure}
\caption{
Up quark distribution $xu_{v}(x)$. The solid line is for attraction in
$u-d$ quark pairs, $D=0.37.$ The dashed line corresponds to the spin
force from color magnetism. The large dot-dashed line corresponds to the
Isgur-Karl quark model result from ref.18 and the smaller one is their
 result evolved from (0.25 GeV$)^{2}$ to 15 GeV$^{2}$. The EMC data are
from ref.7.}
\label{fig5}
\end{figure}
\begin{figure}
\caption{
Ratio of unpolarized neutron to proton structure functions for the
Isgur-Karl model from ref.18 (dotted line) and evolved from
(0.25 GeV$)^{2}$ to 15 GeV$^{2}$ (dot-dashed line). The solid line is
for attraction with $D=0.37.$ The EMC data are from ref.21.}
\label{fig6}
\end{figure}
\begin{figure}
\caption{
Proton asymmetry $A^{p}_{1}(x)$ with data from refs.7 and 25. The
solid line is for attraction between $u-d$ quark pairs, $D=0.37,$ in
the light-cone quark model and the dot-dashed line its evolution from
(0.6 GeV$)^{2}$ to 11 GeV$^{2}$; the dashed line is for the spin force
from color magnetism.}
\label{fig7}
\end{figure}
\newpage
\section*{ References}
\noindent 1.  See, e. g., F. Close, {\it An Introduction to Quarks and
Partons,} Acad. Press (New York, 1979), and references therein.
\par
\medskip
\noindent 2.  A. Manohar and H. Georgi, Nucl. Phys. ${\bf B234}, 189 (1984)$;
 S. Weinberg, Phys. Rev. Lett. {\bf 65}, 1181 (1990).
\par
\medskip
\noindent 3.  P. A. M. Dirac, Rev. Mod. Phys. {\bf 21}, 392 (1949).
\par
\medskip
\noindent 4.  W. Konen and H. J. Weber, Phys. Rev. ${\bf D41}, 2201 (1990);$
Z. Dziembowski, Phys. Rev. ${\bf D37}, 778 (1988)$; I. G. $Az$nauryan,
A. S. Bagdasaryan, and N. L. Ter-Isaakyan, Phys. Lett. ${\bf B112}, 393 (1982)
,$ Sov. J. Nucl. Phys. {\bf 36},743 (1982) [Yad. Fiz. {\bf 36}, 1743 (1982)];
F. Coester, in {\it Nuclear and Particle Physics on the Light Cone,} eds. M.B.
Johnson and L.S. Kisslinger, World Scientific, Singapore (1989).
\par
\medskip
\noindent 5.  H. J. Weber, Phys. Lett. ${\bf B287}, 14 (1992)$, and Ann. Phys.
(N.Y.) {\bf 207}, 417 (1991); I. G. Aznauryan, Preprint CEBAF Th-92-17.
\par
\medskip
\noindent 6.  F. Schlumpf, Preprint SLAC-PUB-6218, to be published in Phys.
Rev {\bf D}.
\par
\medskip
\noindent 7.  J. Ashman {\bf et al.}, EMC  Collab., Phys. Lett. ${\bf B206},
364$ (1988); Nucl. Phys. ${\bf B328}, 1 (1989)$.
\par
\medskip
\noindent 8.  For a review and further refs. see, e.g.{\bf ,} G. Altarelli,
in Proc. "E. Majorana" Summer School, Erice, Italy, ed. A. Zichichi, 1989,
Plenum Press, and R. L. Jaffe and A. Manohar, Nucl. Phys. ${\bf B337}, 509
(1990)$.
\par
\medskip
\noindent 9.  B.-Q. $Ma$, J. Phys. ${\bf G17}, L53 (1991)$; B.-Q. $Ma$ and
Q.-R. Zhang, Univ. Frankfurt Preprint Hep$-ph/9306241.$
\par
\medskip
\noindent 10. H.J. Melosh, Phys. Rev. ${\bf D9}, 1095 (1974)$.
\par
\medskip
\noindent 11. L.A. Kondratyuk and M.V. Terent'ev, Yad. Fiz. {\bf 31}, 1087
(1980) [Sov. J. Nucl. Phys. {\sl 31}, 561 (1980)].
\par
\medskip
\noindent 12. Z. Dziembowski, in Lecture Notes in Physics {\bf 417}, 192
(1992), eds. K. Goeke, P. Kroll and H.-R. Petry, Springer, Berlin.
\noindent 13. A. de Rujula, H. Georgi, and S. Glashow, Phys. Rev.
${\bf D12}, 147 (1975)$.
\par
\medskip
\noindent 14. N. Isgur and G. Karl, Phys. Rev. ${\bf D18}, 4187 (1978)$.
\par
\medskip
\noindent 15. S. Godfrey and  N. Isgur, Phys. Rev. ${\bf D32}, 189 (1985)$;
S. Capstick and N. Isgur, ibid. ${\bf D34}, 2809 (1986)$; S. Capstick,
preprint CEBAF-TH-92-09.
\par
\medskip
\noindent 16. J. D. Bjorken and S. D. Drell, {\it Relativistic Quantum
Mechanics} (McGraw-Hill, New York, 1964).
\par
\medskip
\noindent 17. M. Weyrauch and H. J. Weber, Phys. Lett. ${\bf B171}, 13 (1986)$;
H. J. Weber and H. T. Williams, ibid. ${\bf B205}, 118 (1988)$.
\par
\medskip
\noindent 18. M. Traini, L. Conci and U. Moschella, U. Trento-Louvain preprint
 $(1992)$.
\par
\medskip
\noindent 19. J. Cohen and  H. J. Weber, Phys. Lett. ${\bf B165}, 229 (1985)$.
\par
\medskip
\noindent 20. D. Krupa {\bf et al.}, J. Phys. $G{\bf 10}, 455 (1984)$.
\par
\medskip
\noindent 21. A.C. Benvenuti {\bf et al.}, Phys. Lett. ${\bf B237}, 599
(1990)$.
\par
\medskip
\noindent 22. X. Song and J. $Du$, Phys. Rev. ${\bf D40}, 2177 (1989)$.

\par
\medskip
\noindent 23. J. Aubert {\bf et al.}, Nucl. Phys. ${\bf B259}, 189 (1985)$,
ibid.${\bf B293}, 740 (1987)$; P. Amaudruz {\bf et. al.}, Phys. Lett.${\bf
B295},
 159 (1992)$.
\par
\medskip
\noindent 24. Z. Dziembowski, H.J. Weber, L.Mankiewicz and A. Szczepaniak,
Phys. Rev. ${\bf D 39}, 3257 (1989)$.
\par
\medskip
\noindent 25. G. Baum {\bf et al.}, Phys. Rev. Lett. ${\bf 51}, 1135 (1981)$.
\end{document}